\documentclass[twocolumn,superscriptaddress,amsfont,amssymb,amsmath, showpacs,balancelastpage, nofootinbib]{revtex4-1}

\usepackage{amsmath,amssymb}
\usepackage{graphicx,subfigure}
\usepackage[colorlinks,linkcolor=magenta,anchorcolor=cyan,citecolor=blue]{hyperref}

\def\lf{\left}
\def\rt{\right}
\def\be{\begin{equation}}
\def\ee{\end{equation}}
\def\ba{\begin{eqnarray}}
\def\ea{\end{eqnarray}}

\begin{document}

    \title{Time crystals in primordial perturbations}

    \author{Hao-Hao Li}
	 \affiliation{ School of Physics, University of Chinese Academy of Sciences, Beijing 100049, China}
	 
    \author{ Yun-Song Piao}
    \affiliation{ School of Physics, University of Chinese Academy of Sciences, Beijing 100049, China}
    \affiliation{ Institute of Theoretical Physics, Chinese Academy of Sciences, P.O. Box 2735, Beijing 100190, China}

\begin{abstract}
 Cosmological time crystal (TC) corresponds to a matter state where the periodic motion of field forms a limit cycle in its phase space. We explore what would happen if it existed in inflationary phase. It is found that the limit cycle responsible for TC will
inevitably cause the periodic oscillation of the primordial perturbation spectrum. The oscillatory patterns of the spectrum depend on the TC parameters, and so encode the crystalline
patterns of TC.

\end{abstract}

    \date{\today}
	\maketitle

\textit{Introduction} -- Time crystals (TC) were put forward by Wilczek in 2012
\cite{Wilczek:2012jt}\cite{Shapere:2012nq}, which described the
time-periodic self-organized structures. In this new perspective,
the spontaneous symmetry breaking occurs in the time-direction.
Recently, the concept of TC has developed quickly in physical
community, in particular condensed matter physics. The discrete TC
has been discovered in the condensed matter experiments
\cite{Yao:2017dis}\cite{Zhang:2016obs}.

In Ref.\cite{Bains:2015gpv}, based on the $P(X,\phi)$ theory,
Bains, Hertzberg and Wilczek proposed the cosmological TC, where
$X=-(\nabla\phi)^2$, which corresponds that a limit cycle should
exist in $ \lf(\phi,\dot{\phi} \rt) $ phase space, see FIG.1.
However, such a cycle will inevitably cross the null energy
condition (NEC) curve, which suggests the violation of NEC
\cite{Bains:2015gpv,Easson:2016klq}. It is well-known that the
violation of NEC will cause the instability of perturbation on
short scales. Recently, this instability has be cured by applying
the higher-order derivative operator of $\phi$
\cite{Easson:2018qgr} (equivalently $R^{(3)}\delta g^{00}$, as in
\cite{Cai:2016thi,Creminelli:2016zwa,Cai:2017tku,Cai:2017dyi,Kolevatov:2017voe},
where $R^{(3)}$ is the Ricci scalar on the 3-dimensional spacelike
surface). Is such a stable cosmological TC observable? It is still
an open question.

According to \cite{Bains:2015gpv}, the cosmological TC may appear
in inflationary phase. If so, the time-periodic patterns of TC
might be imprinted in the power spectrum of primordial
perturbation. In our work, we will explore this possibility, and
show that the limit cycle in $ \lf(\phi,\dot{\phi} \rt) $ in phase
space will result in the periodic oscillation of the primordial
spectrum, which so encodes the crystalline patterns of TC, as in
original concept \cite{Shapere:2012nq}. Here, the TC model
considered is a special subclass of G-inflation
\cite{Kobayashi:2010cm}\cite{Burrage:2010cu}.

\textit{A stable cosmological TC} --
In Ref.\cite{Bains:2015gpv}, the cosmological TC was built by
applying such a Langrangian,
    \ba
    {\cal L}={M_P^2\over 2}R+\sum_{n,m\neq 0}c_{n,m}\phi^{2n}\lf(X/2\rt)^{m}-V(\phi) ,
\label{calL}
    \ea
with the coefficient $c_{0,1}=c_{0,2}=1$, $c_{1,1}=\beta>0$,
and the Higgs-like potential \be V\left( \phi
\right)=\Lambda+{1\over 4\alpha} \lf(\alpha\phi^2-1\rt)^2 , \ee
where $\beta>0$ and $\alpha>0$ are the parameters, both impact the
period of cosmological TC \cite{Bains:2015gpv,Easson:2016klq}.
Here, although the Lagrangian (\ref{calL}) is that of k-inflation \cite{ArmendarizPicon:1999rj,Garriga:1999vw}, the inflation is actually potential-driven, and $\Lambda=const.$ is responsible for
the inflation. We have \be M_P^2 H^2=\rho/3=\frac{\left(\beta
\phi^2-1\right)}{6}{\dot \phi}^2+{{\dot\phi}^4\over  4} +
\frac{V}{3}.  \ee

We solve the motive equation of $ \phi $ numerically and plot it in FIG.1. We see that $\phi$ oscillate around the maximum
point of its potential. As showed in Ref.\cite{Bains:2015gpv},
such solutions are the attractors, which form the limit cycles in
$ \lf(\phi,\dot{\phi} \rt) $ phase space. The amplitude and period
of oscillations keep constant, such a oscillatory matter state is called the cosmological TC.
\begin{figure}[htbp]
	\includegraphics[width=3.4in]{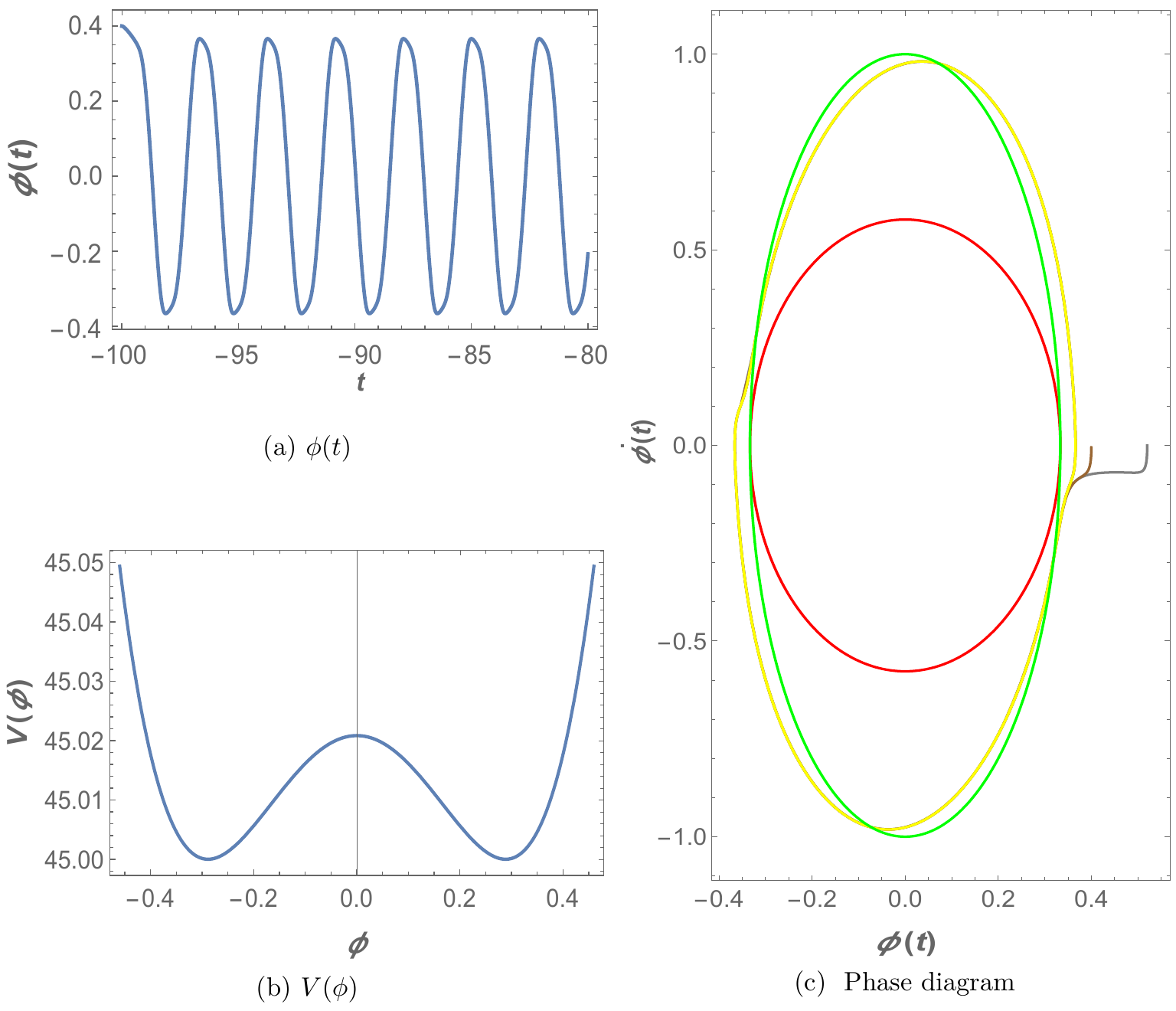}\label{fig1}
	\caption{The evolution of background field $\phi$, its Higgs-like
		potential and the limit cycle. Here, the Planck mass $ M_P\sim10^3 $. 
		In $ \lf(\phi,\dot{\phi} \rt) $ phase space, the yellow one is the
		limit cycle, which is the attractor \cite{Bains:2015gpv}. See the
		gray and brown curves for the trajectories of $\phi$ with different initial $ \lf(\phi_0,\dot{\phi}_0 \rt) $. Inside the green curve, the NEC is violated. Inside the red ellipse, the field is ghost-like. }
\end{figure}

The existence of limit cycles requires that the NEC must be
violated, because the sign of $ \dot{H} $ will change
periodically. The NEC condition for any null vextors $ k^{\mu} $
is $T_{\mu\nu}k^{\mu}k^{\nu}=P_X\left( \nabla_{\mu}\phi k^{\mu}
\right)^2 \geq 0 $. In FIG.1(c), the boundary of NEC violation is an elliptic curve $P_X=0$, which is
$\dot{\phi}^2+\beta\phi^2=1$, where $P=\sum_{n,m\neq
0}c_{n,m}\phi^{2n}\lf(X/2\rt)^{m}$, see the Lagrangian
(\ref{calL}). Another elliptic curve is $\rho_X=2XP_{XX}+P_X=0$, which
is $3\dot{\phi}^2+\beta\phi^2=1$. Inside it, $ \rho_X<0 $.

Because the limit cycle will not cross the ellipse with
$\rho_X=0$, but must cross the boundary of NEC violation, the
bound on the oscillating amplitude of the field is
\cite{Easson:2016klq}
    \begin{gather}
    \frac{1}{\sqrt{\beta}}<\phi_{max} <\sqrt{\frac{1+\sqrt{1+\alpha}}{\alpha}}. \label{width}
    \end{gather}
which shows how the width of the limit cycle depends on the
parameters $ \alpha $ and $ \beta $.



It is well-known that if the NEC is violated, the Langangian
(\ref{calL}) will be instable in perturbative level, since
$c_s^2<0$ is inevitable, $c_s$ is the sound speed of perturbation.
Here, we will cure it by considering Galileon field operator $\sim
\Box\phi$.


  It is simple to work in unitary gauge $ \phi=\phi\left(t \right)$.
  The EFT Lagrangian of cosmological perturbation corresponding to
  the Lagrangian (\ref{calL}) with the Galileon operator $\sim
  \Box\phi$ is
  \begin{equation}
  	\begin{split}
  	\mathcal{L}=&\frac{M_P^2}{2}R-\lambda(t)-c(t)g^{00}\\&+\frac{1}{2}M_2^4(t)\left(\delta
  	g^{00}\right)^2-\frac{1}{2}m_3^3(t)\delta K \delta g^{00}
  	\label{LEFT}
  	\end{split}
  \end{equation}
  where
  \begin{gather}
  \delta g^{00}=1-\frac{X}{\dot{\phi}^2\left(t \right)}
  \end{gather}
  \begin{gather}
  \delta K=-\frac{1}{\left( X\right)^{1/2} }\left[ \Box\phi+\frac{\phi^{\sigma}\phi_{\sigma\gamma}\phi^{\gamma}}{X} \right]-3H
  \end{gather}
  and the coefficients $ c(t)=-M_P^2\dot{H} $, $
  \lambda(t)=M_P^2\left( \dot{H}+3H^2  \right)$ , $M_2^4(t)={\dot
  	\phi}^4P_{XX}$, $m_3^3$ is free, $K$ is the extrinsic curvature on
  the 3-dimensional spacelike surface,  $\Box\phi$ is related to
  $\delta K$ and $ \phi^{\sigma}=\nabla^{\sigma}\phi $, see, e.g.Ref.\cite{Cheung:2007st}.
  
The action of scalar perturbation $\zeta $ for \eqref{LEFT} is
\begin{align}
S_{\zeta}^{(2)}=&\int a^3 Q_s\left[\dot{\zeta}^2-
c_s^2\frac{\left(\partial_i\zeta\right)^2
    }{a^2} \right]d^4x,
    \end{align}
    where \begin{gather}
    Q_s=\frac{M_P^2\left( 8M_P^2{\dot\phi}^4P_{XX}-4M_P^4\dot{H}+3m^6_3\right) }{\left(M_P^2H-m_3^3/2\right)^2}, \label{qs}\\
c_s^2Q_s=d\lf[\frac{2M_P^4a}{\left(2M_P^2H-m_3^3\right)}\rt]/(adt)
-M_P^2. \label{csqs}
    \end{gather}
Here, the stabilities of perturbations require both $Q_s>0$ and
$c_s^2>0$.

According to \eqref{csqs}, without the operator $\delta K \delta
g^{00} $, we have
    \begin{gather}
    c_{s}^2Q_s= -M_P^2{\dot H}/H^2.
    \end{gather}
When the NEC is violated, ${\dot H}>0$. Thus we have else
$c_s^2<0$, or $Q_s<0$, the corresponding evolution is instable.
Considering the operator $\delta K \delta g^{00} $, we have
\begin{equation}
	\begin{split}
	c_{s}^2Q_s=& -{M_P^4{\dot H}\over
		\lf(M_P^2H-m^3_3/2\rt)^2}\\&+{M_P^2\left( M_P^2Hm_3^3/2+{\dot m_3^3}/2-m_3^6/4\right) \over
		\lf(H-m^3_3/2\rt)^2}.
	\end{split}
\end{equation}
Now, if one properly adjusts the coefficient $m_3^3$ of $\delta K
\delta g^{00} $, it is possible that both $Q_s>0$ and $c_s^2>0$
are satisfied simultaneously. Thus the operator $\delta K \delta
g^{00} $ may be applied to cure the instabilities of scalar
perturbations in cosmological TC.

It has been found in
Refs.\cite{Cai:2016thi,Creminelli:2016zwa,Cai:2017tku}, see also
\cite{Cai:2017dyi,Kolevatov:2017voe,Mironov:2018oec}, that the
operator $R^{(3)}\delta g^{00}$ is significant for building
``heathy" nonsingular cosmologies (avoiding No-Go Theorem
\cite{Libanov:2016kfc,Kobayashi:2016xpl}), see also
\cite{Ye:2019frg,Ye:2019sth} for stable nonsingular cosmologies
with the operators $N^{\prime2}$, $N^\prime K$ and $(\partial
N)^2$ in DHOST theory \cite{Langlois:2018dxi}, where $N$ is the
lapse. In Ref.\cite{Easson:2018qgr}, $R^{(3)}\delta g^{00}$ has
been adopted for stable cosmological TC. However, here we only
consider the inflation scenario, which has been proved to be
singular in the past (past-incomplete). Thus the operator $\delta
K \delta g^{00} $ ($\sim \Box\phi$) is enough to dispel the
instabilities, as in G-inflation
\cite{Kobayashi:2010cm,Burrage:2010cu}.

Here, it is convenient to assume that $ m_3^3 $ is proportional to
$ H $, $ m_3^3=xM_P^2 H \left(x\neq0 \right)$. According to \eqref{qs} and \eqref{csqs}, we have
\begin{gather}
    c_s^2=\frac{\left(2x-x^2 \right)}{3x^2}+{\cal O}(\epsilon,
    {{\dot\phi}^4P_{XX}\over H^2}),
    \end{gather}
noting that $ {\dot\phi}^4P_{XX} $ is far smaller than $ H^2 $
during inflation. We keep $c_s^2\approx 1$, so get $ x={1}/{2} $
(which brings $Q_s\approx M_P^2/3>0$). We plot $c_s^2$ in FIG.2. The
blue line is $c_s^2$ without the operator $\delta K \delta g^{00}
$. In contrast, the magenta line is that with $\delta K \delta
g^{00} $. Due to the existence of limit cycle, both $H$ and $\dot
H$ are oscillating. Thus although $c_s^2\approx 1$, it is affected by $\epsilon$ , ${{\dot\phi}^4P_{XX}/ H^2}$ and is also oscillating. In certain sense, the oscillations of both $H$ and $c_s^2$ actually encode the crystal shape of cosmological TC.
\begin{figure}[htbp]
	\subfigure[$\epsilon$]{\includegraphics[width=2.9in]{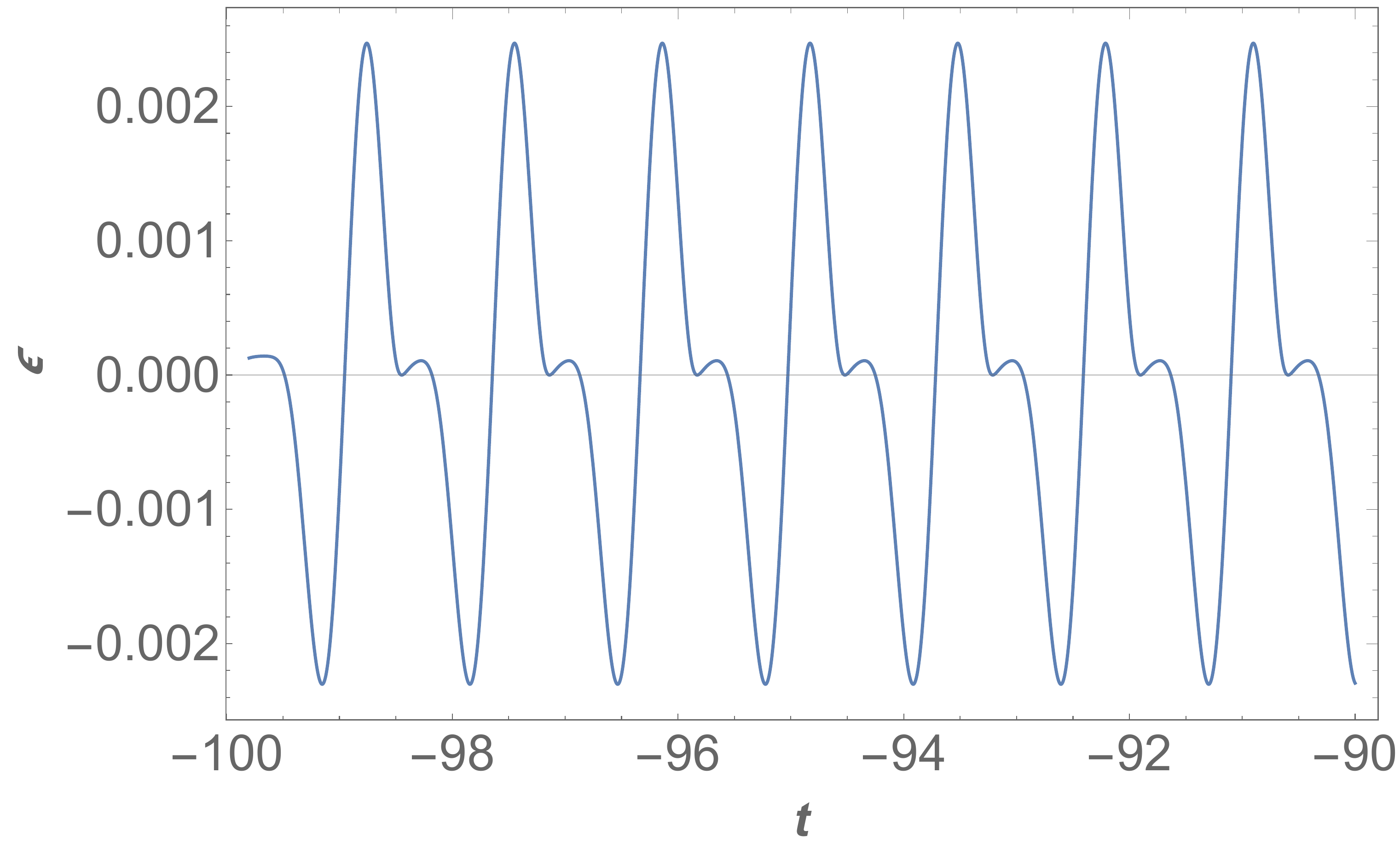}}\label{fig2a}
	\\
	\subfigure[$c_s^2$ ]{\includegraphics[width=2.8in]{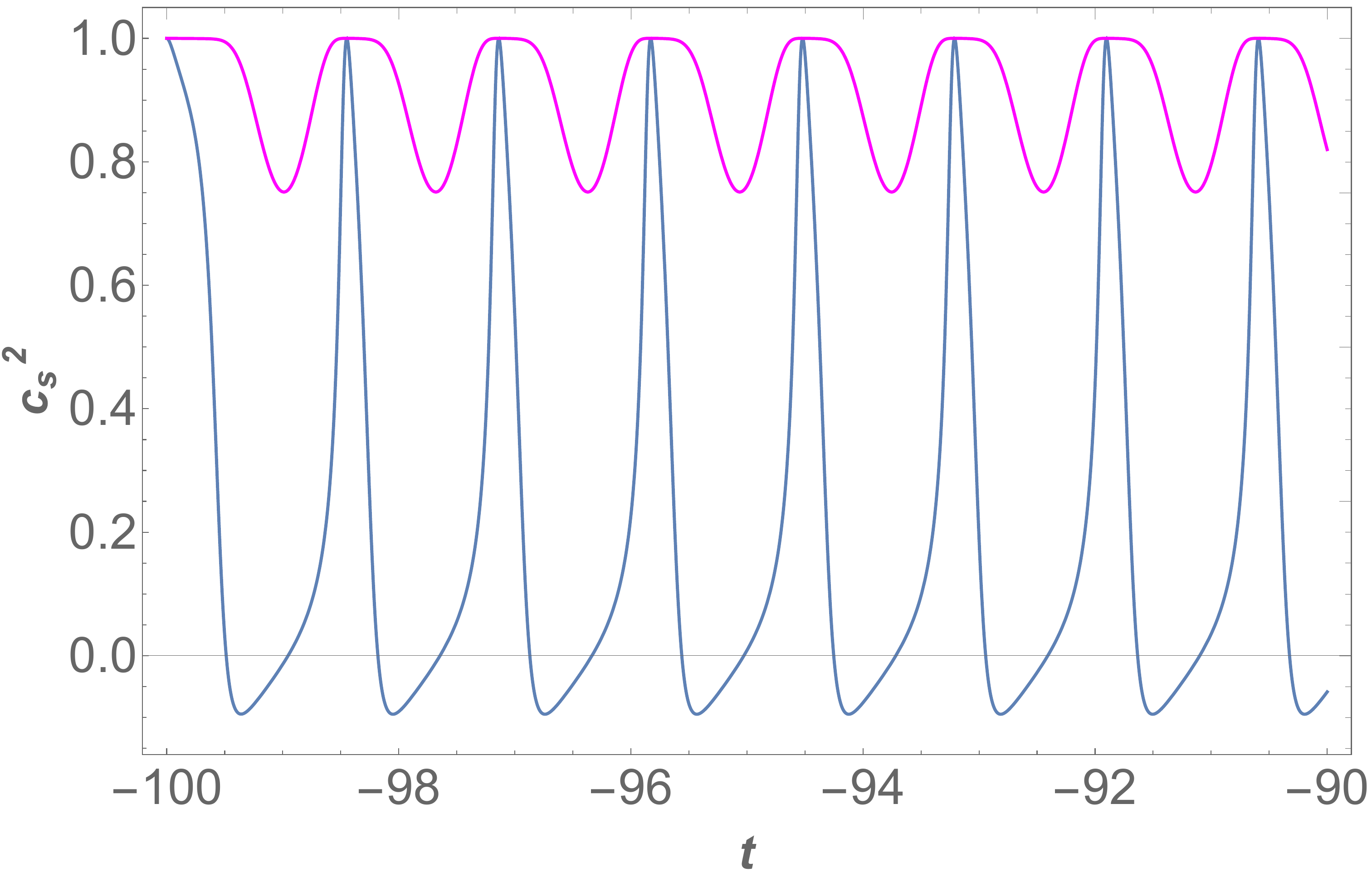}}\label{fig2b}
	\caption{The evolutions of $\epsilon=-{\dot H}/H^2$ and $c_s^2$.
		In (b), the blue line is $c_s^2$ without the Galileon operator $\sim \Box\phi$. }
\end{figure}

\textit{Oscillatory patterns of power spectrum}  -- The power spectrum of primordial perturbation is given by
    \begin{gather}
    P_{\zeta}(k)=\frac{k^3}{2\pi^2}\left|
    \zeta_k\right|^2_{c_sk=aH} . 
    \end{gather}
In the momentum space, the motive equation of $\zeta_k$ is
\begin{gather}
\ddot{\zeta_k}+\left(3H+\frac{\dot{Q_s}}{Q_s}\right)\dot{\zeta_k}+c_s^2\frac{k^2}{a^2}\zeta_k=0. \label{zetak}
\end{gather}
During the inflation, $|\epsilon|=|{\dot H}/H^2|\ll 1$, but
$|{{\dot{Q_s}}\over {Q_sH}}|\ll 1$ is not always valid. Thus we
will solve Eq.(\ref{zetak}) numerically.

Defining the Mukhanov-Sasaki variable $u_k=y\zeta_k $, we rewrite
Eq.(\ref{zetak}) as
    \begin{gather}
    {u_k}''+\left(k^2-\frac{y''}{y}\right)u_k=0,  \label{MSE}
    \end{gather}
where $y=a\sqrt{2Q_sc_s} $, e.g.\cite{Khoury:2008wj}, $
{d}\eta=c_s{d}t/a $ and $ '={{d}}/{{d}\eta}$. This suggests that
the initial state for $u_k$ may be the Bunch-Davies state
$u_k=\frac{e^{-ik\eta}}{\sqrt{2k}}$. In physical time, it equals
to
    \begin{gather}
    \zeta_k=\frac{e^{-ik\int  \frac{c_s}{a}\mathrm{d}t}}{y\sqrt{2k}}, \\
    \dot{\zeta_k}=\left(\frac{-ikc_s}{a}-\frac{\dot{y}}{y}\right)\zeta_k.
    \end{gather}
We plot the perturbation spectrum $P_{\zeta}(k)$ in FIG.3 for different limit cycles. As expected, $P_{\zeta}(k)$ is oscillating
periodically, which is actually an inevitable result of the limit
cycle existing in $ \lf(\phi,\dot{\phi} \rt) $ phase space.

\begin{figure}[htbp]
    \subfigure[ The limit cycles]{\includegraphics[width=2.8in]{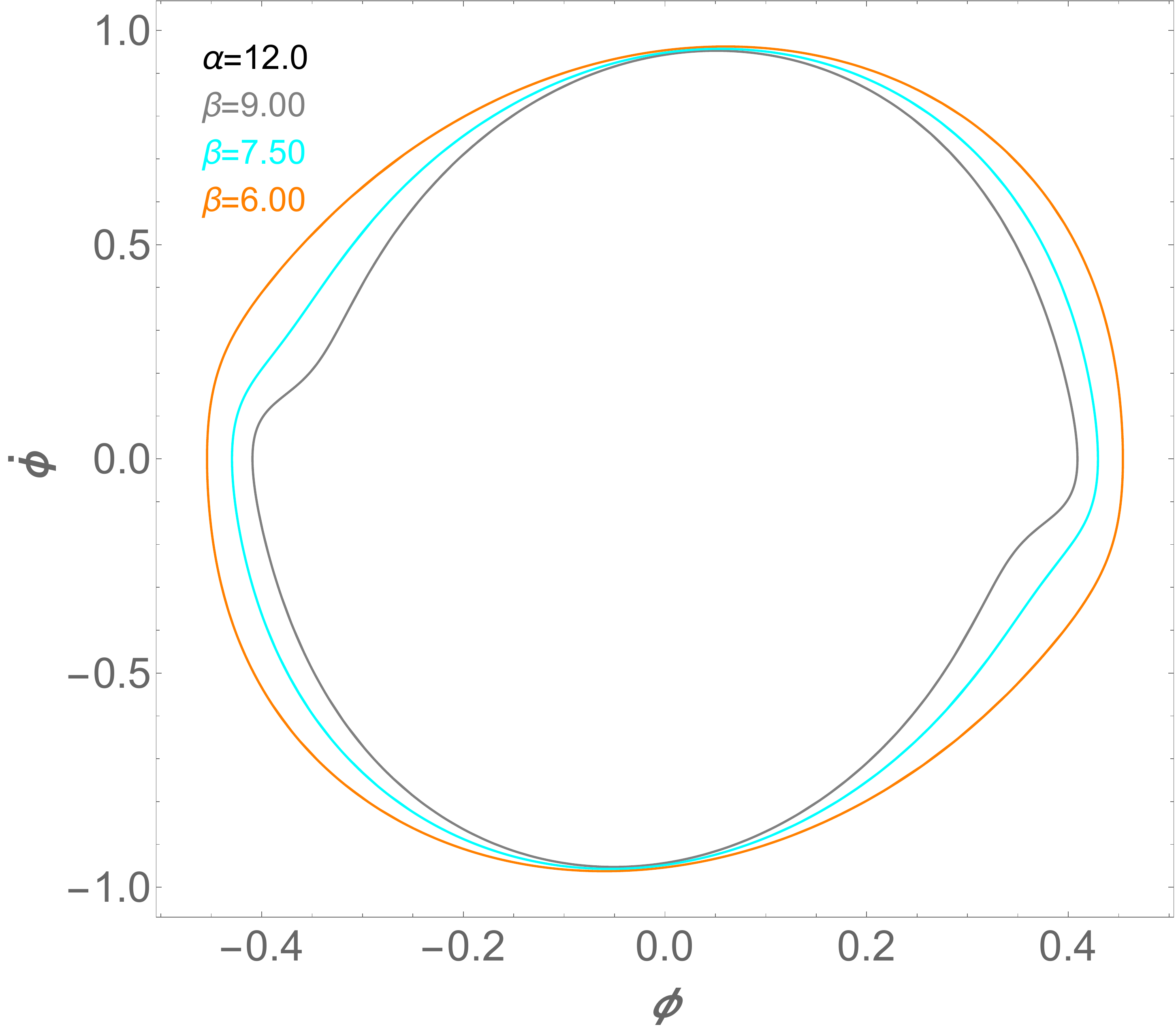}}\label{fig4a}
       \\
    \subfigure[ The corresponding power spectrum]{\includegraphics[width=3in]{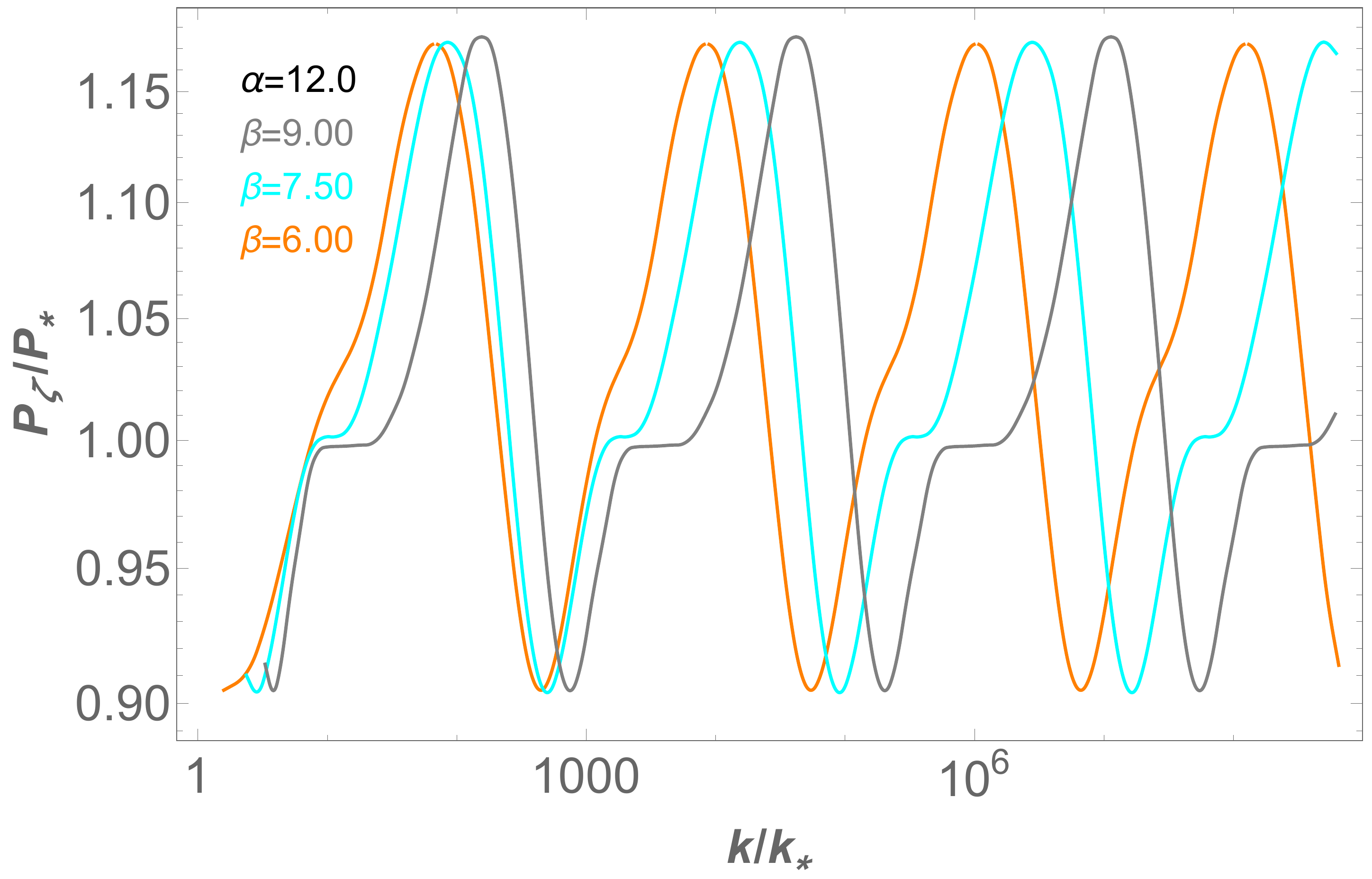}} \label{fig4b}
\caption{The corresponding power spectrum $P_\zeta(k)$ for different limit cycles with $ \Lambda=45. $}
    \end{figure}

It is interesting to search for the possible oscillating patterns of power spectrum $P_{\zeta}(k)$ for different parameters $\alpha$ and $\beta$, which control the shapes of limit cycles. As found in
Ref.\cite{Bains:2015gpv}, the larger $ \beta $ is, the larger allowed range of $\alpha $ is. But a bigger $ \beta $ will bring a shorter minor axis of ghost elliptic curve, see Eq.(\ref{width}). We plot some power spectrums in FIG.4 with $ \beta=30 $ and different values of $\alpha$. We show just four kinds of oscillating patterns. When $ \alpha $ increases, the ``ripple" in the bottom of spectrum will move up, and after crossing the peak, it will descend. When $ \alpha $ is fixed and $ \beta $ is changed, the case is reverse. Thus the shape of limit cycle (equivalently the crystalline patterns of cosmological TC) will be encoded in the different oscillating patterns of power spectrum.

\begin{figure}[htbp]
        \includegraphics[width=1.65in]{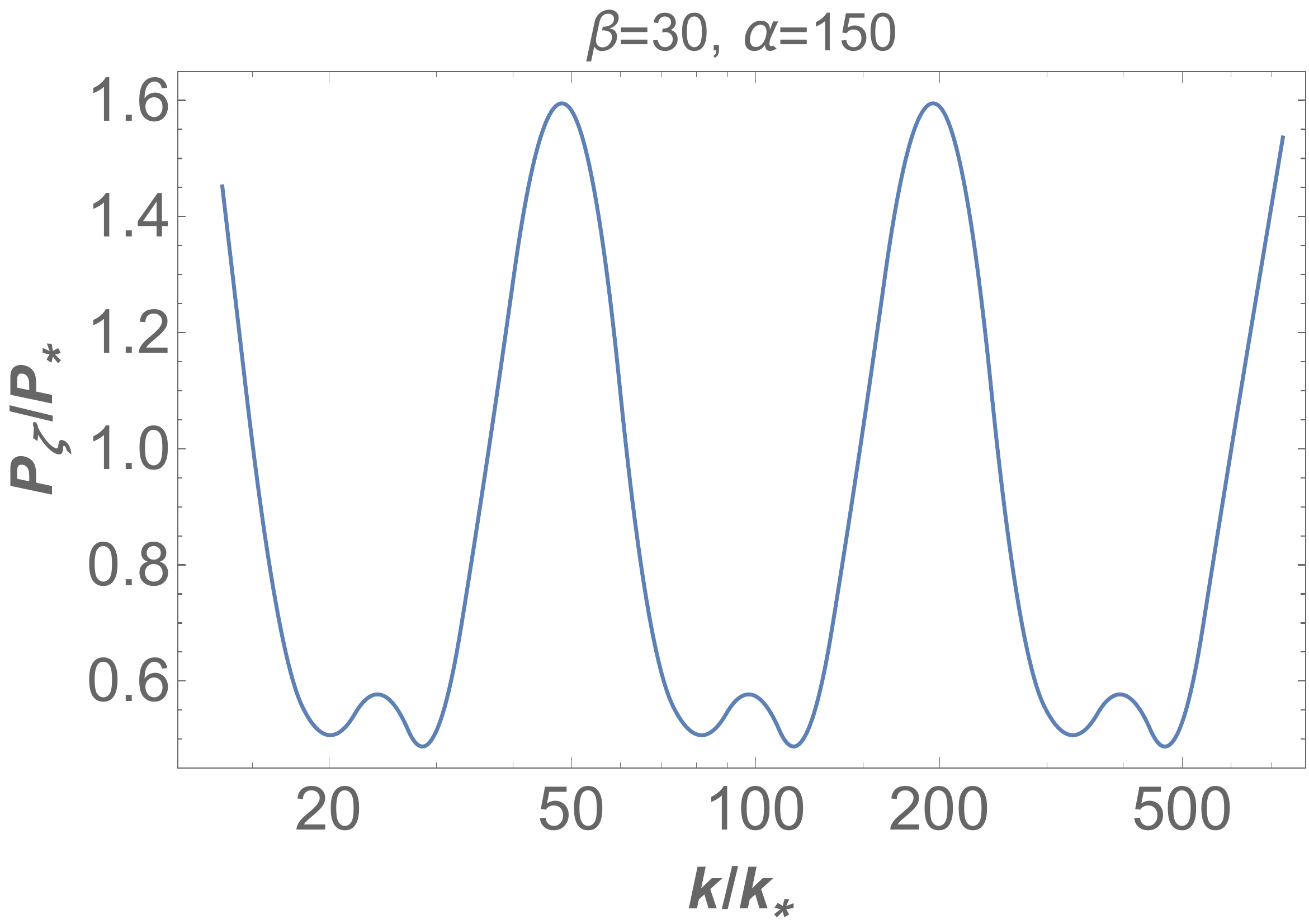}\label{fig5a}
        \,
        \includegraphics[width=1.65in]{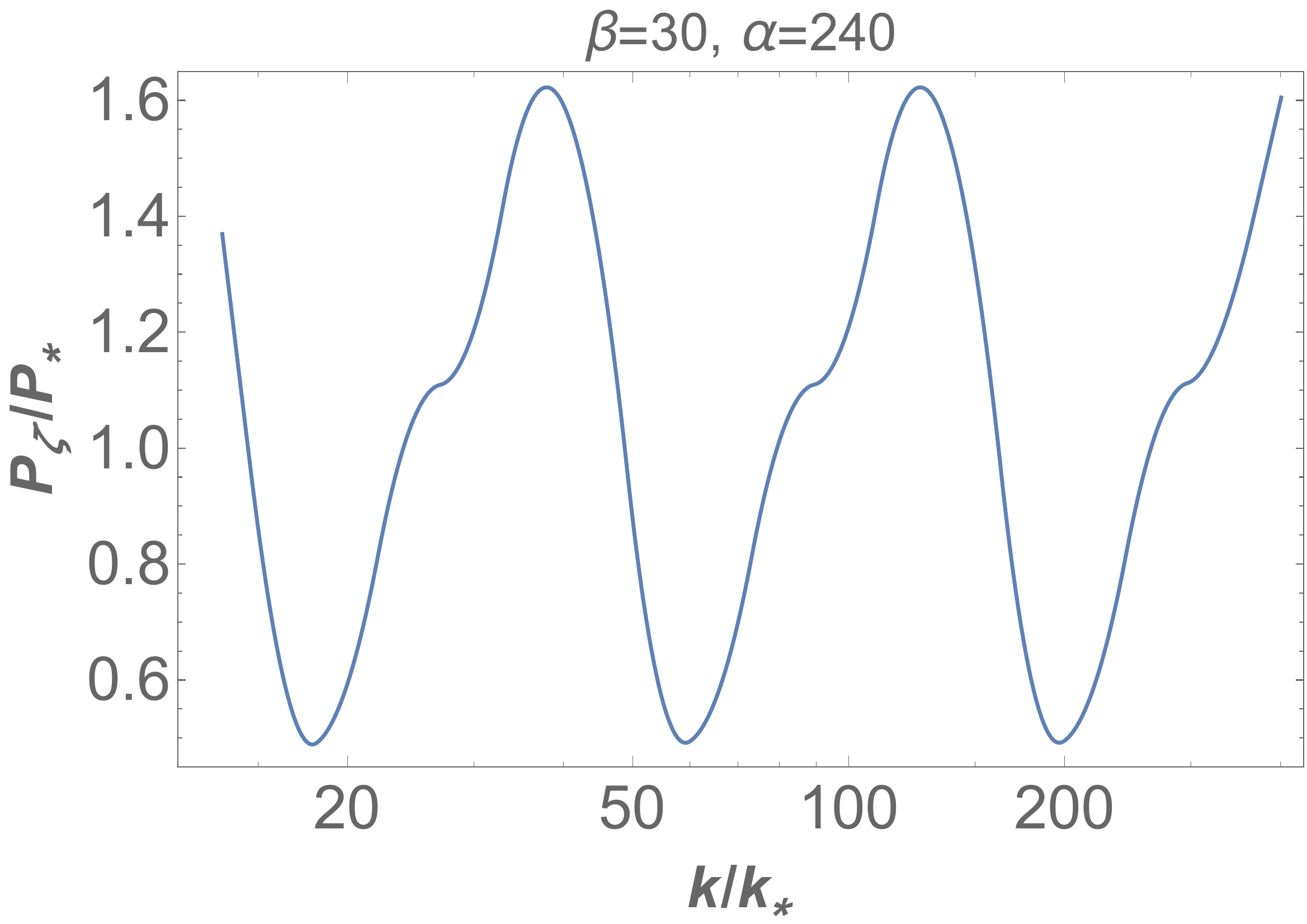}\label{fig5b}
        \\ [12pt]
        \includegraphics[width=1.65in]{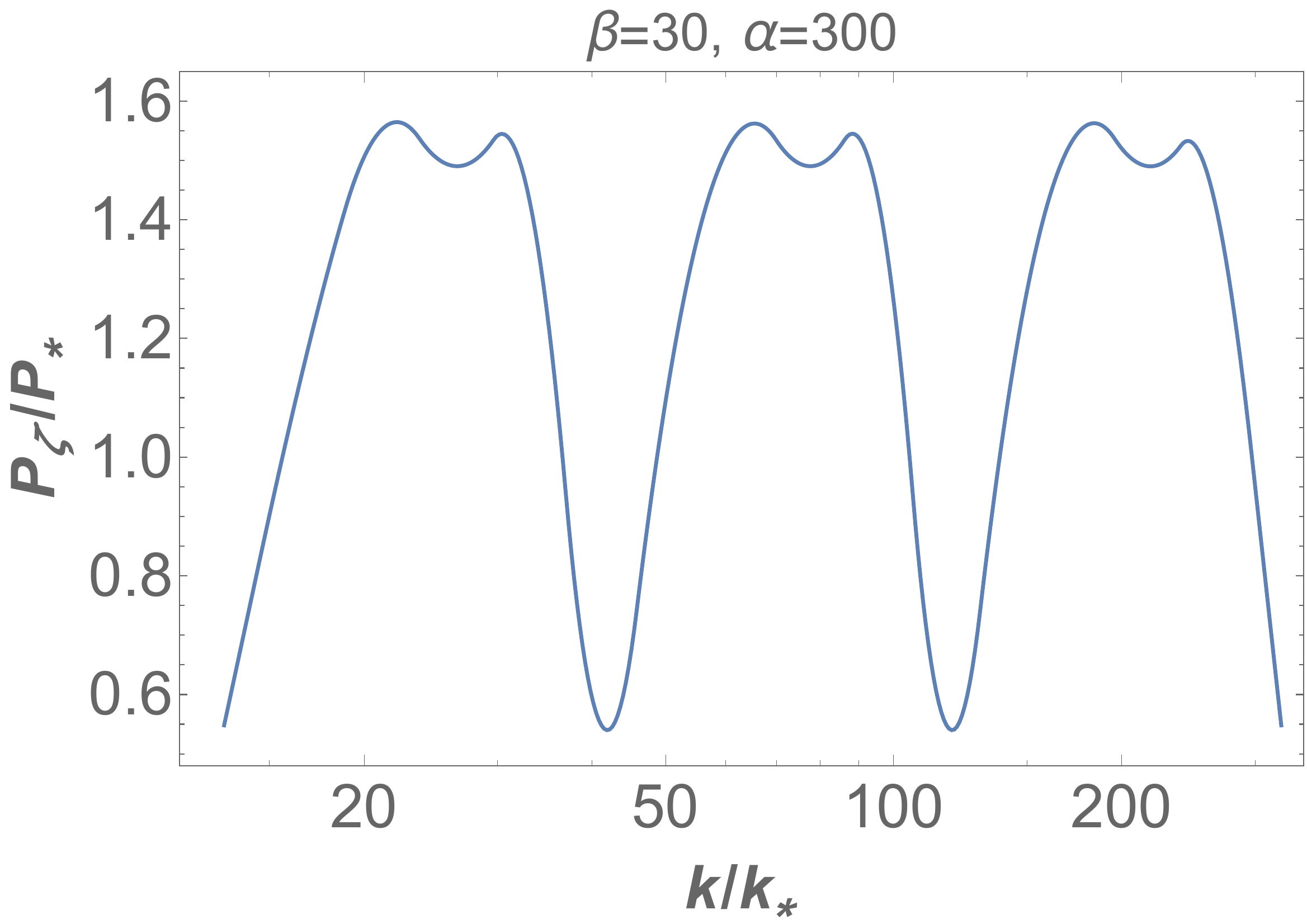}\label{fig5c}
        \,
        \includegraphics[width=1.65in]{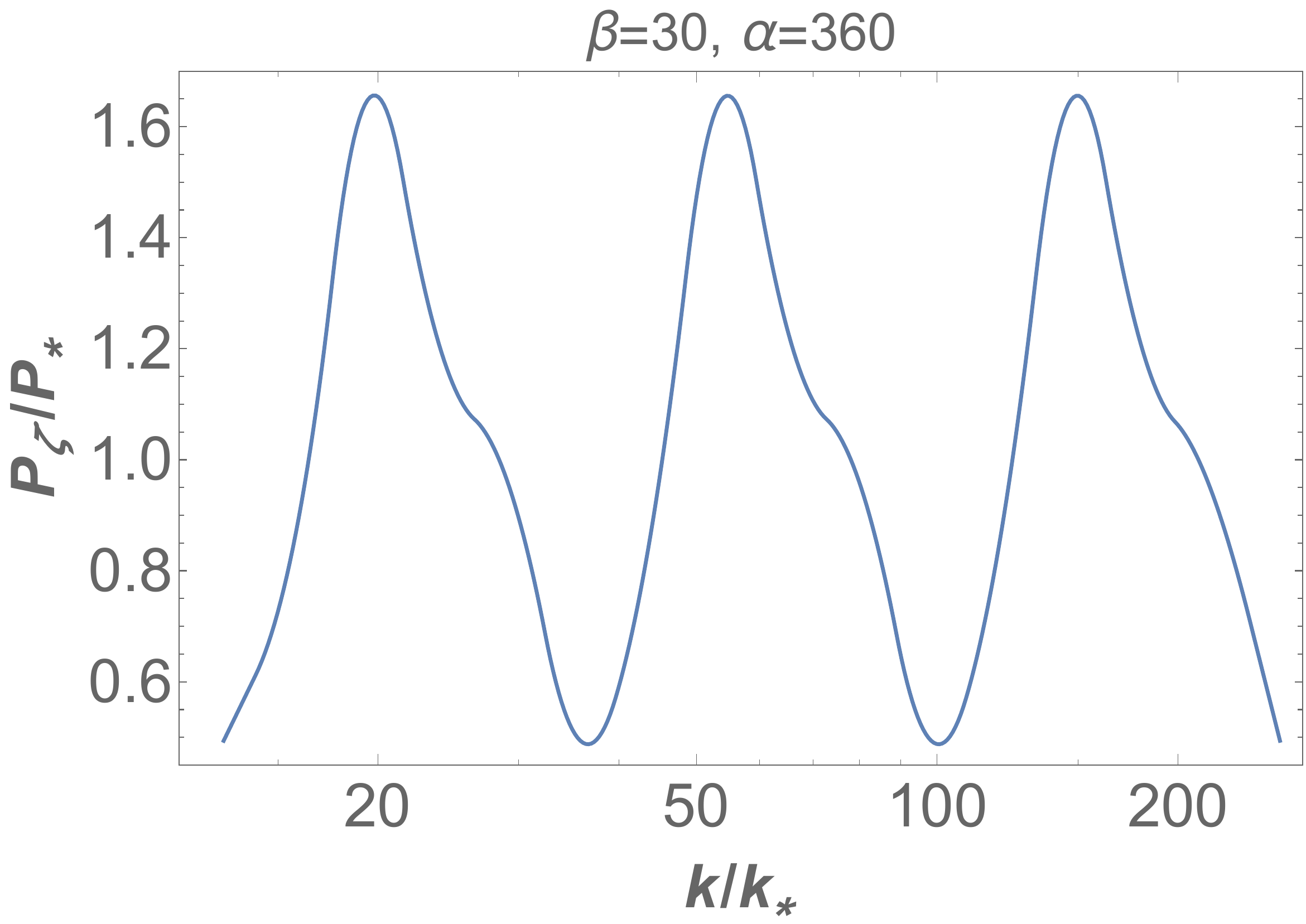}\label{fig5d}
\caption{The possible oscillatory patterns of $P_\zeta(k)$, which
actually encode different crystalline patterns of cosmological TC with $ \Lambda=10. $
}
    \end{figure}

\textit{Discussion} --  Recently, as a special matter state, the cosmological TC has been implemented stably. How to look for such a matter state in
cosmological observations is a significant issue.

In our work, we explored what would happen if the cosmological TC existed in the inflationary phase. We apply the operator $
m_3^3(t)\delta K \delta g^{00} $ ($\sim\Box \phi$) to remove the instability relevant with $c_s^2<0$, and calculated the spectrum of corresponding primordial perturbation. It is found that the limit cycle responsible for TC in $ \lf(\phi,\dot{\phi} \rt) $
phase space will inevitably cause the oscillation of spectrum. We point out that the oscillatory patterns of the primordial spectrum depends on the TC parameters, and so encode the crystalline patterns of TC. Thus if the cosmological TC existed in inflationary phase, the imprint of TC might be preserved in the CMB. It will be interesting to search for the corresponding signals in Planck data.

 There would not be an inflationary stage forever. TC will dissolve due to interactions with a `waterfall' field $ \sigma $\cite{Linde:1993cn}. During inflation phase, field $ \sigma $ is stabilized at its only minimum where $ \sigma=0 $. When $ \phi $ is smaller than the effective mass, the `waterfall' field $ \sigma $ becomes tachyonic and ends inflation. Another field coupled with $ \phi $ may offer a large no-gaussianity which can be observable in the feature experiment. It is a new view to observe cosmological TC. In addition, adding external matter can have very strong implications for stability of models with nonstandard kinetic terms where the perturbations of $ \phi $ near the ghost region $ c_s^2 < 0 $\cite{Easson:2013bda}. However, in our model, the  $ m_3^3(t)\delta K \delta g^{00} $ term will stay the perturbations far away from the ghost region, so it is still a stable model.

It is well-known that the inflation scenario itself can not avoid the initial singularity. The cosmological TC has been also investigated in certain theories of modified gravity
\cite{Das:2018bzx,Feng:2018qnx,Das:2018ylw}, in particular Ref.\cite{Easson:2018qgr} with the operator $R^{(3)}\delta g^{00}$. This might provide a link of the cosmological TC to the nonsingular cosmologies, which is worthy of deeply exploring.

 \section*{Acknowledgments}
We thank Gen Ye and Yong Cai for helpful discussions. This work is
supported by NSFC, Nos.11575188, 11690021.

 \appendix
 \section*{Appendix}
In this paper, we set units $ c=\hbar=1 $
and the sign $(-+++)$.

It is possible to work out the analytic solution of $P_\zeta$, if
we assume that, like the slow-roll parameter $
\epsilon=-\frac{\dot{H}}{H^2} $, other parameters
    \begin{gather}
    \epsilon_s=\frac{\dot{c_s}}{Hc_s}\, , \quad\kappa=\frac{\dot{\epsilon_g}}{H\epsilon_g}, \, \quad \delta=\frac{\dot{\kappa}}{H\kappa} \label{eq28}
    \end{gather}
are also slow-roll, where $\epsilon_g=Q_sc_s$.

Defining the parameter $\epsilon_r=\epsilon+\epsilon_s $, we have
    \begin{gather}
    \frac{\mathrm{d}}{\mathrm{d}\eta}\left(\frac{c_s}{aH} \right)
    =\epsilon_r-1,
    \end{gather}
which gives ${aH\over c_s}\simeq-\frac{1}{\eta}\left(1+\epsilon_r
\right)$. In slow-roll approximation, ${y''}/{y}$ in
Eq.(\ref{MSE}) is
\begin{gather}
\frac{y''}{y}\approx\frac{\nu^2-1/4}{\eta^2},
\end{gather}
where $ \nu=\frac{3}{2}+\epsilon_r+\frac{1}{2}\kappa $. We choose
Bunch-Davies vacuum $u_k=\frac{e^{-ik\eta}}{\sqrt{2k}} $ as the
initial condition of $u_k$ when $ aH\ll c_sk $. On super-Hubble
scales $\left( -k\eta\ll 1\right)  $, the exact solution of
Eq.(\ref{MSE}) is \cite{Stewart:1993bc,Riotto:2002yw}
    \begin{gather}
    u_k=e^{i(\nu-\frac{1}{2})\frac{\pi}{2}}2^{\nu-1}\frac{\Gamma(\nu)}{\sqrt{\pi
    k}}(-k\eta)^{\frac{1}{2}-\nu}.
    \end{gather}
    Thus
    \begin{gather}
    P_{\zeta}(k)=\frac{k^3}{2\pi^2}\left| \frac{u_k}{y} \right|^2_{c_sk=aH}\simeq \frac{2^{2\nu-3}\Gamma^2(\nu)H^2}{2\pi^3\epsilon_gc_s^2\left(1+\epsilon_r
    \right)^{2\nu-1}}.
    \label{PKA}\end{gather}
However, generally, $\epsilon_s$ and $\epsilon_g$ are not slow-roll, so the
slow-roll approximation is not precise. We have checked
numerically that $P_\zeta(k)$ in (\ref{PKA}) is not exactly
similar to the numerical result.



\begin{thebibliography}{99}

 \bibitem{Wilczek:2012jt}
   F.~Wilczek,
   Phys.\ Rev.\ Lett.\  {\bf 109}, 160401(2012)
   [arXiv:1202.2539 [quant-ph]].

 \bibitem{Shapere:2012nq}
   A.~Shapere and F.~Wilczek,
   Phys.\ Rev.\ Lett.\  {\bf 109}, 160402(2012)
   [arXiv:1202.2537 [cond-mat.other]].

 \bibitem{Yao:2017dis}
   N.~Y.~Yao, A.~C.~Potter, I.~D.~Potirniche and A.~Vishwanath,
   Phys.\ Rev.\ Lett.\  {\bf 118}, 030401(2017)
   [arXiv:1608.02589 [cond-mat.dis-nn]].

 \bibitem{Zhang:2016obs}
   J.~Zhang, P.~W.~Hess, A.~Kyprianidis, P.~Becker, A.~Lee, J.~Smith, G.~Pagano, I.~D.~Potirniche, A.~C.~Potter and A.~Vishwanath,
   Nature.\ {\bf 543}, 217-220(2016)
   [arXiv:1609.08684v1  [quant-ph]]

 \bibitem{Bains:2015gpv}
   J.~S.~Bains, M.~P.~Hertzberg and F.~Wilczek,
   JCAP {\bf 1705},  no.05,  011(2017)
   [arXiv:1512.02304 [hep-th]].

 \bibitem{Easson:2016klq}
   D.~A.~Easson and A.~Vikman,
   [arXiv:1607.00996 [gr-qc]].

 \bibitem{Easson:2018qgr}
   D.~A.~Easson and T.~Manton,
   Phys.\ Rev.\ D {\bf 99},  no.4,  043507(2019)
   [arXiv:1802.03693 [hep-th]].

 \bibitem{Cai:2016thi}
   Y.~Cai, Y.~Wan, H.~G.~Li, T.~Qiu and Y.~S.~Piao,
   JHEP {\bf 1701},  090(2017)
   [arXiv:1610.03400 [gr-qc]].

 \bibitem{Cai:2017tku}
   Y.~Cai, H.~G.~Li, T.~Qiu and Y.~S.~Piao,
   Eur.\ Phys.\ J.\ C {\bf 77},  no.6,  369(2017)
   [arXiv:1701.04330 [gr-qc]].

 \bibitem{Creminelli:2016zwa}
   P.~Creminelli, D.~Pirtskhalava, L.~Santoni and E.~Trincherini,
   JCAP {\bf 1611},  no.11,  047(2016)
   [arXiv:1610.04207 [hep-th]].
 
 \bibitem{Cai:2017dyi}
   Y.~Cai and Y.~S.~Piao,
   JHEP {\bf 1709},  027(2017)
   [arXiv:1705.03401 [gr-qc]].

 \bibitem{Kolevatov:2017voe}
   R.~Kolevatov, S.~Mironov, N.~Sukhov and V.~Volkova,
   JCAP {\bf 1708}, no. 08, 038 (2017)
   [arXiv:1705.06626[hep-th]].

 \bibitem{Kobayashi:2010cm}
   T.~Kobayashi, M.~Yamaguchi and J.~Yokoyama,
   Phys.\ Rev.\ Lett.\  {\bf 105}, 231302 (2010)
   [arXiv:1008.0603 [hep-th]].

 \bibitem{Burrage:2010cu}
   C.~Burrage, C.~de Rham, D.~Seery and A.~J.~Tolley,
   JCAP {\bf 1101}, 014 (2011)
   [arXiv:1009.2497 [hep-th]].

 \bibitem{ArmendarizPicon:1999rj}
   C.~Armendariz-Picon, T.~Damour and V.~F.~Mukhanov,
   Phys.\ Lett.\ B {\bf 458}, 209 (1999)
   [hep-th/9904075].

 \bibitem{Garriga:1999vw}
   J.~Garriga and V.~F.~Mukhanov,
   Phys.\ Lett.\ B {\bf 458},  219(1999)
   [hep-th/9904176].

 \bibitem{Cheung:2007st}
   C.~Cheung, P.~Creminelli, A.~L.~Fitzpatrick,  J.~Kaplan and L.~Senatore,
   JHEP {\bf 0803}, 014 (2008)
   [arXiv:0709.0293 [hep-th]].


 \bibitem{Mironov:2018oec}
   S.~Mironov, V.~Rubakov and V.~Volkova,
   JCAP {\bf 1810}, no. 10, 050 (2018)
   [arXiv:1807.08361 [hep-th]].

 \bibitem{Libanov:2016kfc}
   M.~Libanov, S.~Mironov and V.~Rubakov,
   JCAP {\bf 1608}, no. 08, 037 (2016)
   [arXiv:1605.05992 [hep-th]].

 \bibitem{Kobayashi:2016xpl}
   T.~Kobayashi,
   Phys.\ Rev.\ D {\bf 94}, no. 4, 043511 (2016)
   [arXiv:1606.05831 [hep-th]].

 \bibitem{Ye:2019frg}
   G.~Ye and Y.~S.~Piao,
   Commun.\ Theor.\ Phys.\  {\bf 71}, no. 4, 427 (2019)
   [arXiv:1901.02202 [gr-qc]].

 \bibitem{Ye:2019sth}
   G.~Ye and Y.~S.~Piao,
   Phys.\ Rev.\ D {\bf 99}, no. 8, 084019 (2019)
   [arXiv:1901.08283 [gr-qc]].

 \bibitem{Langlois:2018dxi}
   D.~Langlois,
   Int.\ J.\ Mod.\ Phys.\ D {\bf 28}, no. 05, 1942006 (2019)
   [arXiv:1811.06271 [gr-qc]].

 \bibitem{Khoury:2008wj}
   J.~Khoury and F.~Piazza,
   JCAP {\bf 0907},  026(2009) [arXiv:0811.3633 [hep-th]].

 \bibitem{Linde:1993cn} 
 A.~D.~Linde,
 Phys.\ Rev.\ D {\bf 49}, 748 (1994)
 [astro-ph/9307002].
 
 \bibitem{Easson:2013bda} 
 D.~A.~Easson, I.~Sawicki and A.~Vikman,
 JCAP {\bf 1307}, 014 (2013)
 [arXiv:1304.3903 [hep-th]].

 \bibitem{Das:2018bzx}
 P.~Das, S.~Pan, S.~Ghosh and P.~Pal,
 Phys.\ Rev.\ D {\bf 98} (2018) no.2,  024004
 [arXiv:1801.07970 [hep-th]].

 \bibitem{Feng:2018qnx}
   X.~H.~Feng, H.~Huang, S.~L.~Li, H.~Lü and H.~Wei,
   arXiv:1807.01720 [hep-th].

 \bibitem{Das:2018ylw}
   P.~Das, S.~Pan and S.~Ghosh,
   Phys.\ Lett.\ B {\bf 791},  66(2019)
   [arXiv:1810.06606 [hep-th]].

 \bibitem{Stewart:1993bc}
   E.~D.~Stewart and D.~H.~Lyth,
   Phys.\ Lett.\ B {\bf 302},  171(1993)
   [gr-qc/9302019].

 \bibitem{Riotto:2002yw}
   A.~Riotto,
   ICTP Lect.\ Notes Ser.\  {\bf 14},  317(2003)
   [hep-ph/0210162].

    \end{thebibliography}
\end{document}